\documentclass[manuscript]{aastex62}
\usepackage{hyperref}
\usepackage{amsmath}
\usepackage{amssymb}
\usepackage{graphicx}
\usepackage{epstopdf}

\shorttitle{high-mass cluster forming region, G25.82--0.17}

\begin{document}

\newcommand{\sour}{G25.82--0.17}
\newcommand{\thmet}{CH$_3$OH 22$_{4}$--21$_{5}$ E}
\newcommand{\masmeto}{CH$_3$OH 8$_{-1}$--7$_{0}$ E}
\newcommand{\sio}{SiO 5--4}
\newcommand{\waterms}{H$_{2}$O 6$_{16}$-5$_{23}$}
\newcommand{\water}{H$_2$O}
\newcommand{\met}{CH$_3$OH}
\newcommand{\mjypb}{mJy beam$^{-1}$}
\newcommand{\inten}{mJy beam$^{-1}$ km s$^{-1}$}
\newcommand{\vel}{km s$^{-1}$}

\title{Multiple outflows in the high-mass cluster forming region, G25.82--0.17}

\author{Jungha Kim}
\affiliation{Department of Astronomical Science, SOKENDAI (The Graduate University for Advanced Studies), 2-21-1 Osawa, Mitaka, Tokyo 181-8588, Japan}
\affiliation{National Astronomical Observatory of Japan, 2-21-1 Osawa, Mitaka, Tokyo 181-8588, Japan }
\author{Mi Kyoung Kim}
\affiliation{Mizusawa VLBI Observatory, National Astronomical Observatory of Japan, 2-21 Hoshi-ga-oka, Mizusawa-ku, Oshu, Iwate 023-0861, Japan}
\affiliation{Otsuma women's university, Faculty of Home Economics, Department of Child Studies, 12 Sanban-cho, Chiyoda-ku, Tokyo 102-8357, Japan}
\author{Tomoya Hirota}
\affiliation{National Astronomical Observatory of Japan, 2-21-1 Osawa, Mitaka, Tokyo 181-8588, Japan }
\author{Kee-Tae Kim}
\affiliation{Korea Astronomy and Space Science Institute, 776 Daedeokdaero, Yuseong, Daejeon 34055, Republic of Korea}
\affiliation{University of Science and Technology, 217 Gajeong-ro, Yuseong-gu, Daejeon 34113, Republic of Korea}
\author{Koichiro Sugiyama}
\affiliation{National Astronomical Observatory of Japan, 2-21-1 Osawa, Mitaka, Tokyo 181-8588, Japan }
\affiliation{National Astronomical Research Institute of Thailand, 260 M.4, T. Donkaew, Amphur Maerim, Chiang Mai 50180, Thailand}
\author{Mareki Honma}
\affiliation{Department of Astronomical Science, SOKENDAI (The Graduate University for Advanced Studies), 2-21-1 Osawa, Mitaka, Tokyo 181-8588, Japan}
\affiliation{Mizusawa VLBI Observatory, National Astronomical Observatory of Japan, 2-21 Hoshi-ga-oka, Mizusawa-ku, Oshu, Iwate 023-0861, Japan}
\author{Do-young Byun}
\affiliation{Korea Astronomy and Space Science Institute, 776 Daedeokdaero, Yuseong, Daejeon 34055, Republic of Korea}
\affiliation{University of Science and Technology, 217 Gajeong-ro, Yuseong-gu, Daejeon 34113, Republic of Korea}
\author{Chungsik Oh}
\affiliation{Korea Astronomy and Space Science Institute, 776 Daedeokdaero, Yuseong, Daejeon 34055, Republic of Korea}
\author{Kazuhito Motogi}
\affiliation{Graduate School of Sciences and Technology for Innovation, Yamaguchi University, Yoshida 1677-1, Yamaguchi 753-8512, Japan}
\author{Jihyun Kang}
\affiliation{Korea Astronomy and Space Science Institute, 776 Daedeokdaero, Yuseong, Daejeon 34055, Republic of Korea}
\author{Jeongsook Kim}
\affiliation{Korea Astronomy and Space Science Institute, 776 Daedeokdaero, Yuseong, Daejeon 34055, Republic of Korea}
\author{Tie Liu}
\affiliation{Shanghai Astronomical Observatory, Chinese Academy of Sciences, 80 Nandan Road, Shanghai 200030, People's Republic of China}
\author{Bo Hu}
\affiliation{School of Astronomy and Space Science, Nanjing University, 22 Hankou Road, Nanjing 210093, People's Republic of China}
\author{Ross A. Burns}
\affiliation{National Astronomical Observatory of Japan, 2-21-1 Osawa, Mitaka, Tokyo 181-8588, Japan }
\affiliation{Korea Astronomy and Space Science Institute, 776 Daedeokdaero, Yuseong, Daejeon 34055, Republic of Korea}
\author{James O. Chibueze}
\affiliation{Space Research Unit, Physics Department, North West University, Potchefstroom 2520, South Africa}
\affiliation{Department of Physics and Astronomy, Faculty of Physical Sciences, University of Nigeria, Carver Building, 1 University Road, Nsukka, Nigeria}
\author{Naoko Matsumoto}
\affiliation{National Astronomical Observatory of Japan, 2-21-1 Osawa, Mitaka, Tokyo 181-8588, Japan }
\affiliation{The Research Institute for Time Studies, Yamaguchi University, Yoshida 167701, Yamaguchi, Yamaguchi 753-8511, Japan}
\author{Kazuyoshi Sunada}
\affiliation{Mizusawa VLBI Observatory, National Astronomical Observatory of Japan, 2-21 Hoshi-ga-oka, Mizusawa-ku, Oshu, Iwate 023-0861, Japan}
\clearpage
	
\begin{abstract}
We present results of continuum and spectral line observations with ALMA and 22 GHz water (H$_2$O) maser observations using KaVA and VERA toward a high-mass star-forming region, G25.82--0.17. 
Multiple 1.3 mm continuum sources are revealed, indicating the presence of young stellar objects (YSOs) at different evolutionary stages, namely an ultra-compact H\textsc{ii} region, G25.82--E, a high-mass young stellar object (HM-YSO), G25.82--W1, and starless cores, G25.82--W2 and G25.82--W3.
Two SiO outflows, at N--S and SE--NW orientations, are identified. 
The CH$_3$OH 8$_{-1}$--7$_{0}$ E line, known to be a class I CH$_3$OH maser at 229 GHz is also detected showing a mixture of thermal and maser emission. 
Moreover, the H$_2$O masers are distributed in a region $\sim0.25\arcsec$ shifted from G25.82--W1.
The CH$_3$OH 22$_{4}$--21$_{5}$ E line shows a compact ring-like structure at the position of G25.82--W1 with a velocity gradient, indicating a rotating disk or envelope. 
Assuming Keplerian rotation, the dynamical mass of G25.82--W1 is estimated to be $>$25 M$_{\odot}$ and the total mass of 20 M$_\odot$-84 M$_\odot$ is derived from the 1.3 mm continuum emission. 
The driving source of the N--S SiO outflow is G25.82--W1 while that of the SE--NW SiO outflow is uncertain. 
Detection of multiple high-mass starless$/$protostellar cores and candidates without low-mass cores implies that HM-YSOs could form in individual high-mass cores as predicted by the turbulent core accretion model. 
If this is the case, the high-mass star formation process in G25.82 would be consistent with a scaled-up version of low-mass star formation. 
\end{abstract}

\section{INTRODUCTION}
The formation of high-mass stars remains incompletely understood despite its large impact on astronomy \citep{zinnecker07,tan14}. 
Even the basic mechanism of high-mass star formation is still controversial. 
However, different theoretical models agree that high-mass young stellar objects (HM-YSOs) require higher mass accretion rates of 10$^{-4}-10^{-3}$ M$_{\odot}$ $\textrm{yr}^{-1}$ \citep{zinnecker07} compared to that of low-mass YSOs \citep[$\sim$10$^{-6}$ M$_{\odot}$ $\textrm{yr}^{-1}$;][]{evans09} to gather mass to the central star by overcoming strong radiation pressure within shorter evolutionary timescales than low-mass cases.
According to theoretical calculations, mass accretion is halted when the central protostellar mass reaches around 8 M$_{\odot}$ due to stellar feedback from the newly-formed central star. 
Non-spherical accretion has been proposed as a possible solution to solve the stellar feedback problem to form OB-type stars \citep{nakano89,jijina96}.
Theoretical studies in recent years show agreement with disk-mediated accretion scenarios for both theoretical models \citep[e.g.,][]{krumholz09,kuiper10,kuiper11,klassen16,beltran18}.
Velocity gradients indicating circumstellar disks around HM-YSOs have been detected with scales of 100--1000 au with the Atacama Large Millimeter/submillimeter Array (ALMA) \citep[e.g.,][]{beltran16, beuther17, cesaroni17}. 
Observational studies of disk--outflow systems are important because outflows could play a significant role in extracting angular momentum and hence, control the mass accretion processes in those systems \citep[e.g.,][]{hirota17}. 
However, detailed observational studies on HM-YSOs are challenging due to large source distances, crowded fields caused by clustering, and short evolutionary timescales.
As such, the number of detailed studies with high spatial/spectral resolutions is still insufficient.

One powerful tool to investigate high-mass star formation is molecular maser emission.
Molecular masers such as water (\water), methanol (\met), and hydroxyl (OH), are known to be associated with a large number of HM-YSOs, and they provide information on the distribution and kinematics of shocked gas their vicinity. 
\met\ masers are divided into two classes based on their pumping mechanisms \citep{menten91}.
Class I \met\ masers (e.g. 44 GHz) are detected offset from the YSOs and are collisionally excited in the shocked regions of extended lobes of low-velocity outflows \citep{kurtz04, cyganowski09}. 
In contrast, class II \met\ masers (e.g. 6.7 GHz) are radiatively pumped and are mostly found in the immediate vicinity of HM-YSOs \citep[e.g.,][]{breen13,fujisawa14}.
They are often associated with rotating and possibly accreting disks \citep[e.g.,][]{bartkiewicz09,sugiyama14,motogi17} or low-velocity outflows traced by infrared emission \citep{debuizer12}. 
In some cases, the class I \met\ masers and the 6.7 GHz class II \met\ masers show spatial overlap \citep[e.g., in DR21(OH)][]{fish11}.
The 22 GHz \water\ masers trace shocked gas associated with various kinds of dynamical structures including low- (a few 10 \vel) and high- ($>$100 \vel) velocity outflows, disks, and H\textsc{ii} regions \citep[e.g.,][]{moscadelli05, motogi16}. 
All aforementioned maser species can be investigated together to trace different dynamical structures around HM-YSOs. 
Moreover, maser features have high intensities and compact structures detectable with very long baseline interferometry (VLBI) observations at milli-arcsecond (mas) scale, which allow measurement of their proper motions.
Studying a large sample of HM-YSOs via their maser emission will help to reveal the roles of disks and outflows in the formation of high-mass stars. 

With this in mind, we initiated a large observing program (LP) using the KVN (Korean VLBI Network) and VERA (VLBI Exploration of Radio Astrometry) array (KaVA), beginning in 2014 \citep{kimj18}.
The KaVA LP aims to establish an evolutionary sequence for the high-mass star formation process by using different maser species and to study the three-dimensional (3D) velocity structures of low- and high-velocity outflows from HM-YSOs \citep[e.g.,][]{torrelles97,moscadelli00,goddi05,goddi11,burns17}.
In total, 87 sources with strong maser emission in one or more maser transitions were selected as an initial sample based on previous single-dish surveys \citep[e.g.,][]{kang15,kim18}.

\sour\ is one of the high-mass star forming regions in our KaVA sample. 
This region is weak or dark at infrared wavelengths \citep[e.g., \textit{Wide-field Infrared Survey Explorer} (WISE);][]{wright10}. 
However, a large scale $^{13}$CO outflow has been reported with the James Clerk Maxwell Telescope (JCMT), indicating that \sour\ is a potential site of high-mass star formation \citep{devillier14}.
In addition, 6.7 GHz \met\ masers were detected in the region with the Very Large Array \citep[VLA; ][]{hu16}, strongly suggesting the existence of a HM-YSO close to the \met\ maser site as the 6.7 GHz \met\ maser is exclusively associated with HM-YSOs \citep{breen13}.
Thus, \sour\ is a suitable target for a detailed study of the HM-YSO(s) associated with outflow and possibly disk through high resolution maser observations.
The systemic velocity of \sour\ is 93.7 \vel\ from ammonia (the NH$_3$ (1,1) line) observations using the Effelsberg 100 m telescope \citep{wienen12}, with a corresponding kinematic distance of 5 $\pm$ 0.3 kpc \citep{green11}.
Hereafter, we will refer to our target, \sour, as G25.82 for simplicity.

In Section 2, our observations of G25.82 are summarized. 
Results and discussions are presented in Sections 3 and 4, respectively.
In Section 5, we summarize this work. 

\section{OBSERVATIONS AND DATA ANALYSIS}
\subsection{ALMA Observations}

ALMA observations of G25.82 in Band 6 (230 GHz) were performed as a part of the KaVA LP project on August 19, and 22, 2016 (project 2015.1.01288.S: P.I. M.-K. Kim).
The number of 12 m antennas was 36 and baseline lengths ranged from 15 to 1462 m. 
The phase tracking center for observations was ($\alpha, \delta$)=(18$^h$39$^m$03$^s$.63, -06{\degr}24{\arcmin}09.{\arcsec}5) in the J2000.0 epoch. 
We set five spectral windows (spws) to cover the \sio\ line, \met\ lines, and continuum emission. 
The 36 GHz series of the \masmeto\ maser line \citep[class I;][]{slysh02}, is included. 
Frequencies of the observed lines are listed in Table 1.
The spectral resolution and bandwidth were 1953.04 kHz ($\sim$2.6 \vel) and 937.5 MHz ($\sim$1300 \vel), respectively.
In addition to the maser transitions, thermal \met\ lines such as \thmet\ were also included.
The on-source time was 33 minutes.
Initial data calibration was carried out using the ALMA pipeline from the CASA\footnote{https://casa.nrao.edu/} package version 4.7 \citep{mcmullin07}. 
The half-power beamwidth (HPBW) of the primary beam was approximately 27$\arcsec$ at the observing frequencies and the synthesized beam size of the continuum data was $0.27\arcsec\ \times 0.24\arcsec$ with a position angle of -83.6. 
Synthesized beams for the line images range from $0.25\arcsec\ \times 0.26\arcsec$ to $0.30\arcsec\ \times 0.26\arcsec$ using a robust weighting of 0.5. 
Plotting and analysis were done with the \textit{Miriad}\footnote{http://www.atnf.csiro.au/computing/software/miriad/} software package \citep{sault95}.

\subsection{KaVA Observations}
We carried out VLBI observations on November 17, 2016, with KaVA toward G25.82 at K-band (22 GHz). 
The position of the phase tracking center of the target was ($\alpha, \delta$)=(18$^h$39$^m$03$^s$.63, -6{\degr}24{\arcmin}09{\arcsec}.5) in the J2000.0 epoch. 
An extragalactic radio continuum source, NRAO 530 was employed for bandpass and delay calibration.
The total bandwidth was 256 MHz (16 MHz $\times$ 16 IFs) and the left-hand circular polarization at a 1 Gbps sampling rate was recorded. 
We analyzed only one of the 16 MHz IF channels containing to the \waterms\ transition.
The spectral resolution was 15.625 kHz ($\sim$0.21 km s $^{-1}$) for the \water\ maser line.
The correlation process was carried out at the Korean-Japan Correlation Center, Daejeon, Korea \citep[KJCC:][]{lee15}.

KaVA data calibration was carried out using the Astronomical Image Processing System (AIPS) developed by National Radio Astronomy Observatory (NRAO)\footnote{http://www.aips.nrao.edu/index.shtml} \citep{vanmoorsel96}.
First, amplitudes were calibrated by using the AIPS task APCAL using measured system temperatures. 
Next, delays and phase offsets were removed by running AIPS task FRING using NRAO530.
Bandpass response was also calibrated using NRAO530. 
Fringe fitting was done on a reference maser component in G25.82 having a local standard of rest (LSR) velocity of 80.5 \vel. 
Imaging and CLEAN (deconvolution) were performed using the AIPS task IMAGR.
Hereafter, a maser `spot' refers to an individual maser emission peak in a spectral channel while a maser `feature' denotes a group of maser spots considered to exist within the same maser cloudlet closely located in space and in consecutive velocity channels.
The peak positions of maser spots were derived by Gaussian fitting to the individual channel images using the AIPS task SAD. 
We employed image pixels greater than 5$\sigma$ noise level for the Gaussian fitting to distinguish maser spots from side lobes with 1$\sigma$ noise level of 23 \mjypb. 

\subsection{VERA Observations}
G25.82 was also observed with VERA at K-band on November 24, 2016, as a part of a monitoring survey toward high-mass star-forming regions using the same phase tracking center as in our KaVA observations.
The dual-beam phase-referencing mode was used for these observations \citep{honma08}, in which the target source and a phase calibrator were observed simultaneously. 
This method is used to derive accurate absolute positions of the \water\ maser spots. 
Correlation processing was done using the Mizusawa software correlator at the National Astronomical Observatory of Japan (NAOJ). 
One IF channel was assigned to G25.82 while 15 IF channels were used for phase reference source, J1846-0651. 
The bandwidth of each IF and spectral resolution settings were the same as KaVA observations, 16 MHz and 15.625 kHz (or 0.21 \vel), respectively. 
We also used image pixels with maser fluxes greater than 5$\sigma$ noise level to identify maser spots with 1$\sigma$ noise level of 24 \mjypb.

Calibration processes were also done with AIPS. 
Dual-beam phase calibration and delay correction were applied to the visibility data \citep{honma08}. 
Instrumental delays and phase offsets were removed by executing AIPS task FRING on DA406 (J1613+3412).
Bandpass calibration was also conducted using DA406.
Finally, residual phases were calibrated by running AIPS task FRING on J1846-0651 and the solutions were applied to G25.82.
Imaging and identifying maser spots was done through the same processes for KaVA data.                        
The \water\ maser spot at a velocity of 80.5 \vel\ was used as a reference of absolute position for calculating the relative positions of other maser spots detected in KaVA observation.

\section{RESULTS}
                                                                                                                                                                                                                                                                                                                                                                                                     \subsection{Continuum Emission} 
\begin{figure}
\centering
\includegraphics{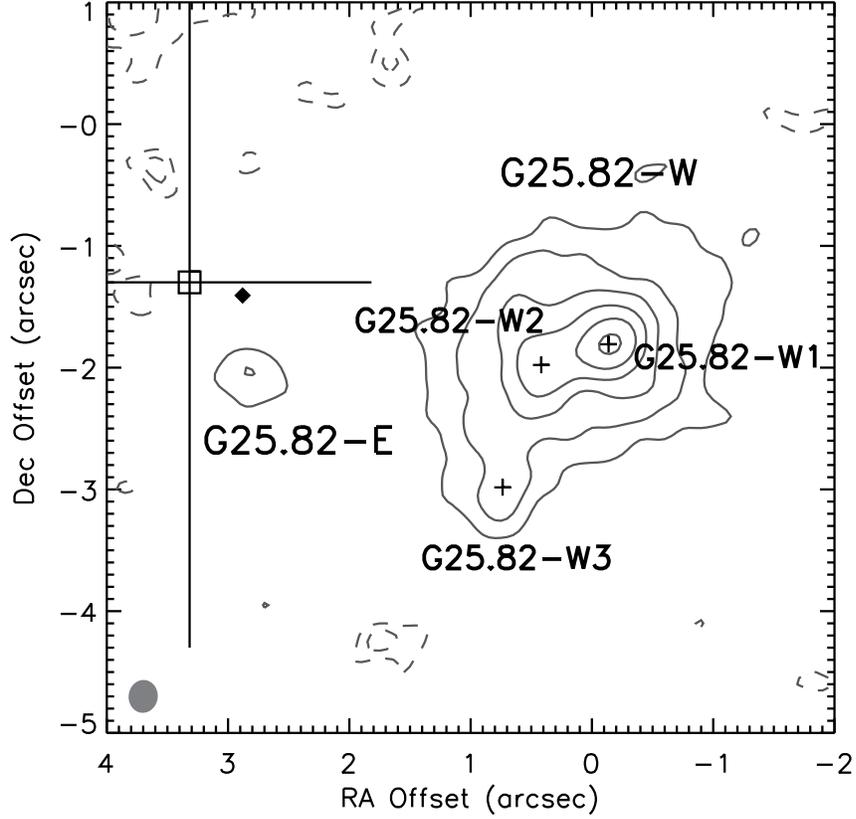}
\caption{Continuum emission map toward G25.82 at 1.3 mm obtained with ALMA.
Contour levels are 10$\sigma$, 20$\sigma$, 40$\sigma$, 80$\sigma$, 160$\sigma$, and 320$\sigma$, where 1$\sigma$ is 0.4 \mjypb\ while dashed contours are negative levels of -1$\sigma$ and -2$\sigma$.
The filled gray circle in the bottom left corner indicates the synthesized beam size of the 1.3 mm continuum image. 
The absolute coordinate of the (0,0) position is (18$^h$39$^m$03$^s$.63, -6{\degr}24{\arcmin}09.5{\arcsec}) in the J2000.0 epoch.
Small crosses denote the positions of 1.3 mm continuum sources in G25.82--W.
An open square indicates the peak position of VLA radio continuum emission with the angular resolution denoted by the cross \citep{hu16}.
In addition, the position of WISE source is marked with the filled diamond.
The absolute position error of ALMA Cycle 3 observations is less than 0.01\arcsec\ while the VLA error is 0.3\arcsec\ \citep{hu16} .}
\label{dust-cont}
\end{figure}

\autoref{dust-cont} shows the distribution of 1.3 mm continuum emission obtained with ALMA in G25.82.
The structure is comprised of two parts (G25.82--E and G25.82--W) with a separation in the E-W direction of 3$\arcsec$. 
The weaker component, G25.82--E, is located 0.6$\arcsec$ south from the peak of WISE 22 $\micron$ emission.
The positional accuracy of the WISE source is indicated by the extent of the diamond in \autoref{dust-cont} (($\Delta\alpha$, $\Delta\delta$)$=$(0.12$\arcsec$, 0.13$\arcsec$)).

Continuum emission at 1.3 mm toward G25.82--W shows a cometary structure and evidence of multiplicity.
Three components were identified by two-dimensional Gaussian fits to emission at intensities larger than 10$\sigma$ of the noise level (1$\sigma=0.4$ \mjypb).
These three components were named as G25.82--W1, G25.82--W2, and G25.82--W3 (plus marks in \autoref{dust-cont}).
The key parameters of these 1.3 mm continuum sources are listed in Table 2.

\subsection{Molecular Line Emission}
A number of molecular lines are detected toward G25.82.
Most of the detected lines are distributed in the vicinity of G25.82--W and show a compact morphology.
In this study, we focus on four molecular lines as representative tracers of high-density gas; the \sio, \masmeto, and \thmet\ lines from the ALMA observations and the \waterms\ maser from the KaVA observations.

\begin{figure}
\begin{center}
\includegraphics{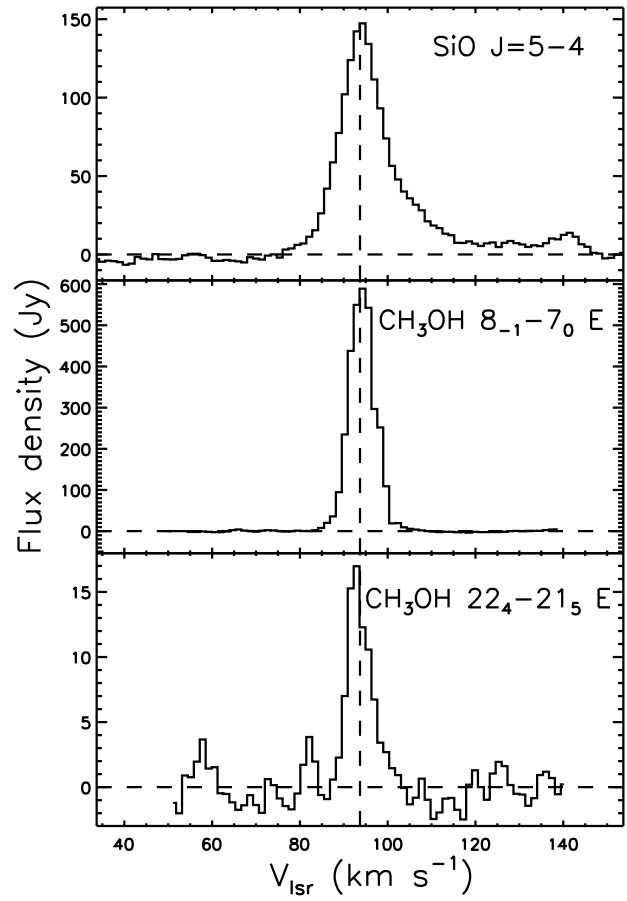}
\caption{Spectra of analyzed molecular transitions toward G25.82 observed with ALMA.
The vertical and horizontal dashed lines represent the systemic velocity \citep[93.7 \vel;][]{wienen12} of G25.82 and the zero baselines, respectively.
Spectra are integrated over the rectangular region ($15\arcsec \times 15\arcsec$) as indicated in \autoref{sio-outflow}. }
\label{alma-spec}
\end{center}
\end{figure}

The profiles of molecular lines obtained with ALMA are shown in \autoref{alma-spec}.
Using a Gaussian profile, robust fitting was done to derive line parameters such as the peak flux density, central velocity, and full width half maximum (FWHM) of the Gaussian.
The central velocities of the \sio, \masmeto, and \thmet\ lines are 94.3 \vel, 94.3 \vel, and 92.9 \vel, respectively.
These are consistent with the adopted systemic velocity of G25.82 (93.7 \vel).
The FWHM of the \masmeto\ and \thmet\ lines are 6.8 \vel\ and 6.3 \vel, respectively.
The \sio\ line has a broader FWHM ($\Delta v \sim$ 11.7 \vel) than the other two lines.

\subsubsection{\sio }
One of the well-known outflow tracers \citep[e.g.,][]{lopez11,codella13,leurini14}, SiO, delineates the spatial and velocity structures of outflowing gas in G25.82.
The outflow components have wing-like spectral features broader than the velocity width of the \thmet\ line (FWHM=6.3 \vel). 
The velocity ranges of the blue- and red-shifted components are defined as from 59.2 \vel\ to 80.1 \vel\ and from 103.7 \vel\ to 157.7 \vel, respectively, for the \sio\ line.  
The red-shifted emission has a larger velocity range by a factor of 2.6 than that of the blue-shifted one.

\begin{figure}
\begin{center}
\plotone{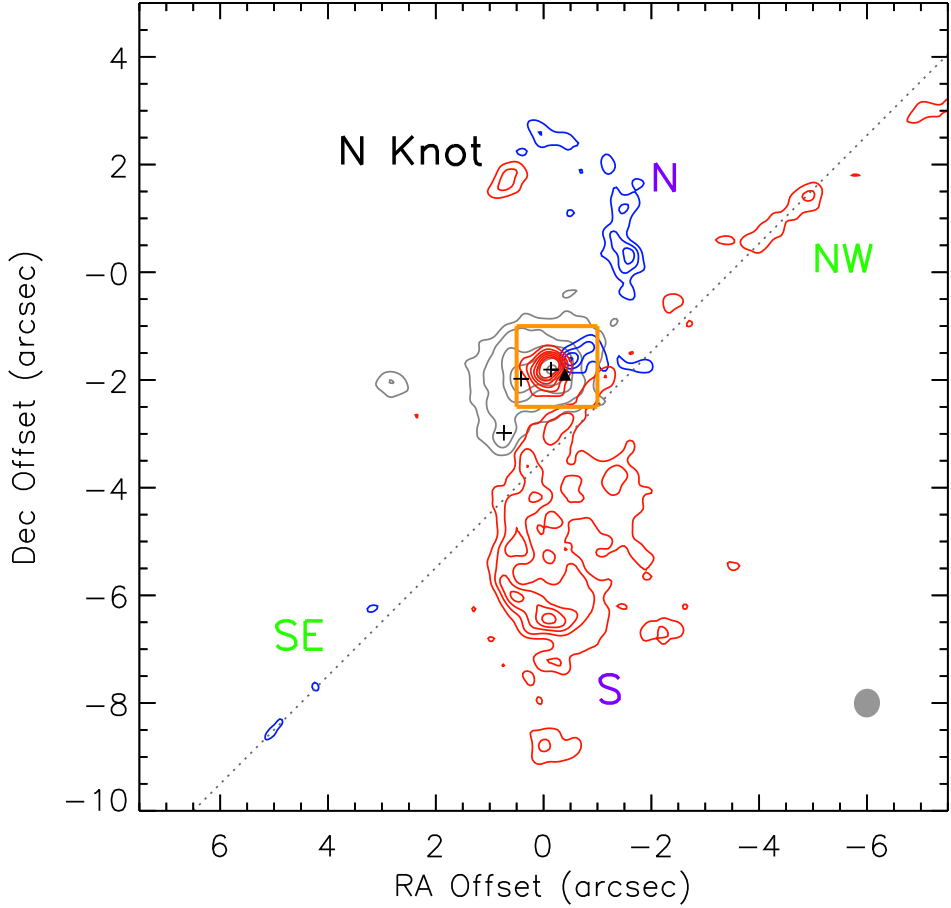}
\caption{Integrated intensities of blue- and redshifted emission of the \sio\ line overlaid onto the 1.3 mm continuum emission (gray contours; see \autoref{dust-cont}).
Blue contours indicate intensities integrated over the velocity range from 59.2 \vel\ to 80.1 \vel\ while red contours represent that from 103.7 \vel\ to 157.7 \vel.
Contours range from 5$\sigma$ to 30$\sigma$ with the intervals of 5$\sigma$, where 1$\sigma$ for blue and red components are 9.8 \inten\ and 18 \inten, respectively. 
Positions of G25.82--W1, W2, and W3 are marked with black crosses while that of the \water\ maser spot at 80.5 \vel\ is denoted by a black triangle. 
The dotted line indicates the outflow axis of SE--NW SiO outflow. 
The orange box indicates the area shown in \autoref{sio-water}.
The filled gray circle in the right bottom corner denotes the synthesized beam size of the SiO map.
The absolute positional uncertainties of water maser features are in all cases $<$ 1 mas.
In addition, that of ALMA Cycle 3 observations is less than 0.01\arcsec.}
\label{sio-outflow}
\end{center}
\end{figure}

\autoref{sio-outflow} shows the integrated intensity maps of the \sio\ line. 
There are at least two SiO outflows in G25.82; the north--south (N--S) outflow and the southeast--northwest (SE--NW) outflow, both of which have red-shifted dominant structures.
The N--S SiO outflow originates from a position $\sim$0.2\arcsec\ northwest of G25.82--W1 and has an extent of about 12\arcsec.
Its total velocity range is from 52.9 \vel\ to 157.7 \vel.
Blue-shifted and redshifted emission exhibit knotted, curved structure, extending to the north and south, respectively.
The SiO emission of this SE-NW outflow is very weak but still significant ($>5\sigma$). 
The blueshifted (SE) and redshifted (NW) components can be seen in the integrated intensity map with the velocity ranges of 75.4-80.1 \vel\ and 103.7-146.9 \vel, respectively.
The SE--NW SiO outflow originates from a position separated at least $\sim$0.8\arcsec\ from the three continuum sources and extends about 17\arcsec.
Further analysis of other outflow tracers such as the CO lines will be helpful to investigate more detailed nature of the SE-NW SiO outflow.
In addition to these bipolar structures, one red-shifted condensation having a high intensity of 12$\sigma$ can be seen in the velocity range from 103 \vel\ to 113 \vel\ at the north of G25.82--W, which is labeled as N Knot in \autoref{sio-outflow}.

\subsubsection{\masmeto}
\begin{figure}
\begin{center}
\includegraphics{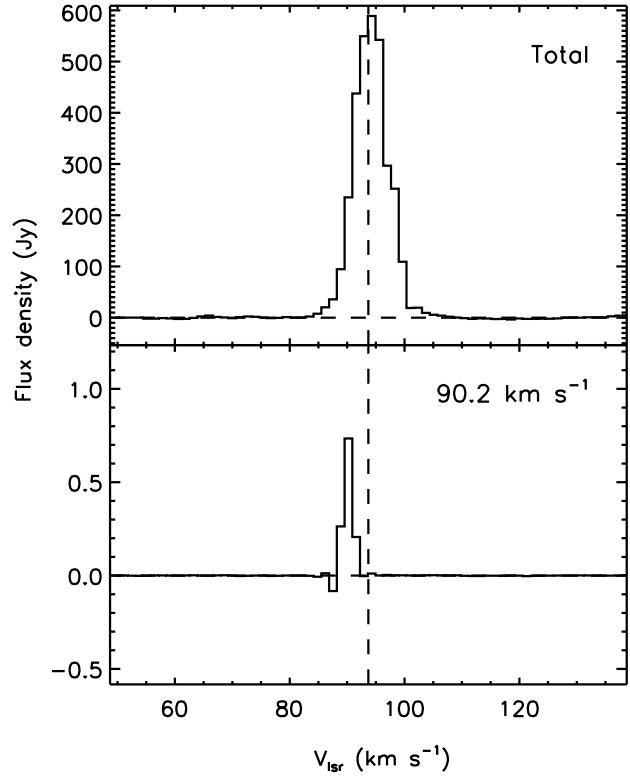}
\caption{Spectra of the \masmeto\ line integrated over whole emitting region (\textit{top}) and a compact component at the central velocity of 90.2 \vel\ (\textit{bottom}).
The vertical dashed lines and the horizontal dashed lines represent the systemic velocity \citep[93.7 \vel;][]{wienen12} of G25.82 and the zero baseline, respectively.}
\label{met-spec}
\end{center}
\end{figure}

\begin{figure*}
\begin{center}
\includegraphics[angle=-90,scale=1.0]{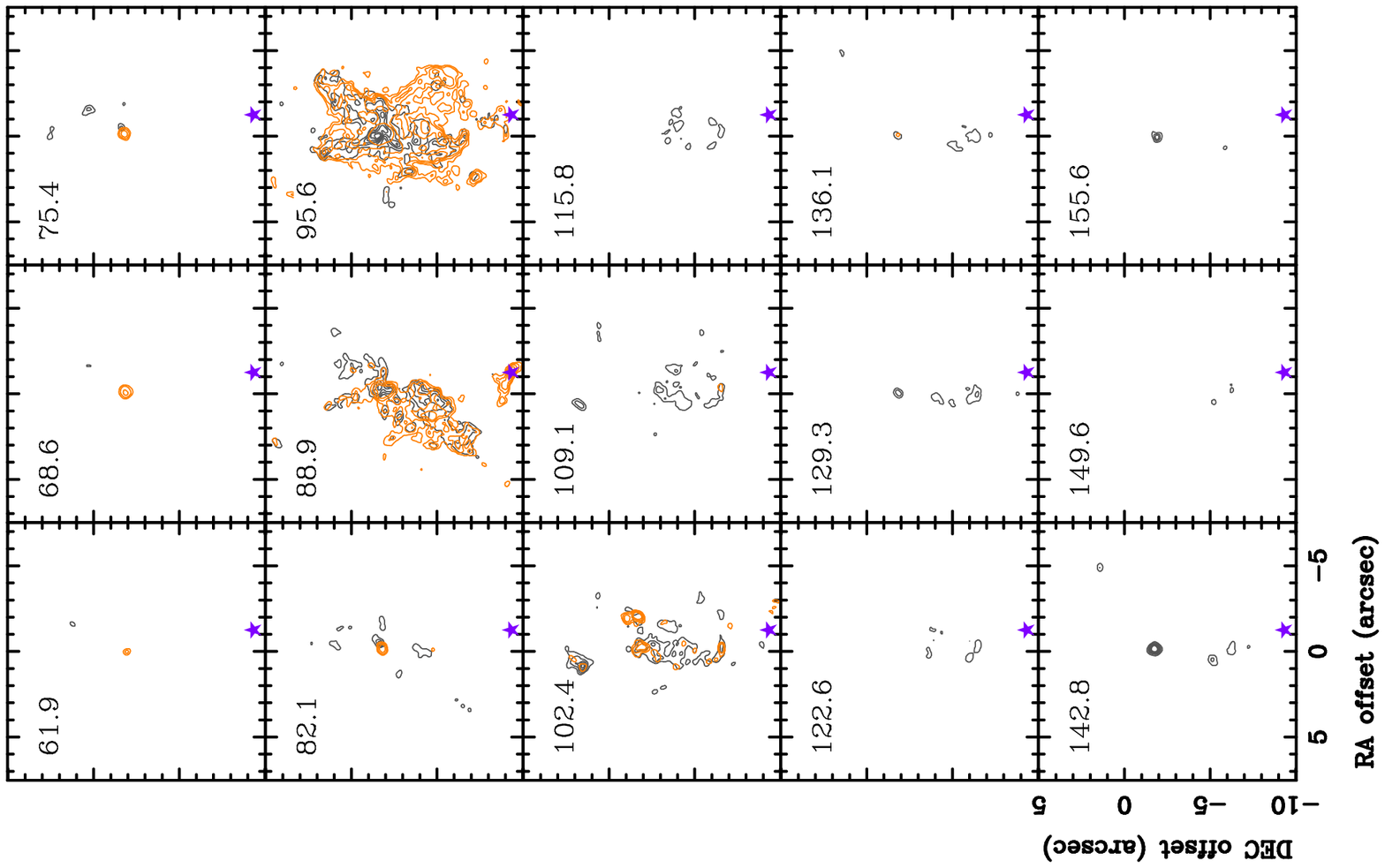}
\caption{Distribution of integrated intensities of the \sio\ (gray contours) and the \masmeto\ (orange contours) lines.
Gray contours start from 5$\sigma$ to 30$\sigma$ with the increasing interval of 5$\sigma$, where 1$\sigma$ is 1.4 \mjypb. 
Orange contours indicate 5$\sigma$, 20$\sigma$, and 80$\sigma$, where 1$\sigma$ is 1.4 \mjypb. 
The compact features seen in orange contours at (0\arcsec, -2\arcsec) at the velocities from 61.9 \vel\ to 82.1 \vel\ and 136.1 \vel\ are contamination from the other molecular lines close to the frequencies of the \sio\ line.
Numbers in the upper left of each panel are respective velocities in \vel.
A purple star denotes the position of the compact structure shown in the bottom panel of \autoref{met-spec}.}
\label{sio-ch3oh}
\end{center}
\end{figure*}

The \met\ line at 229 GHz, \masmeto, is also detected toward G25.82. 
This line is predicted to be a a class I methanol maser \citep{slysh02}. 
Nevertheless, the line profile of the \masmeto\ line emission in G25.82 shows a mixture of maser and thermal emission (\autoref{met-spec}). 
The line width of the \masmeto\ spectrum integrated over the whole G25.82 region is comparable to that of the \thmet\ line emission (upper panel in \autoref{met-spec}). 
Weak wing emission is also found in the profile of the \masmeto\ line.
Compact components of the \masmeto\ line are identified by Gaussian fits to the image.
The spectrum of a spatially compact feature at a central velocity of 90.2 \vel, which is marked with a cross in \autoref{sio-ch3oh}, is shown in the lower panel of \autoref{met-spec}.
It shows a narrower line width ($\sim$2 \vel) compared to that of the total spectrum.
Due to insufficient spatial and spectral resolution, we could derive only the lower limit to the brightness temperature of $>$240~K using the peak flux of the compact feature.

The distribution of the \masmeto\ line is presented in \autoref{sio-ch3oh} (orange contours).
Its extended structure ($\sim$15\arcsec) appears similar to the \sio\ line emission in channels close to the systemic velocity of 93.7 \vel. 
In addition to the overall extended structure, there are compact features of \masmeto\ emission with a size of 1\arcsec\ or less.
These class I \met\ maser features appear to trace lower velocity outflowing gas than that traced by the SiO emission. 
It suggests that the class I \met\ maser at 229 GHz are likely tracing weaker or older shocks compared to the SiO emission \citep[e.g.,][]{yanagida14}.

\subsubsection{\water\ Masers}
\begin{figure}
\begin{center}
\plotone{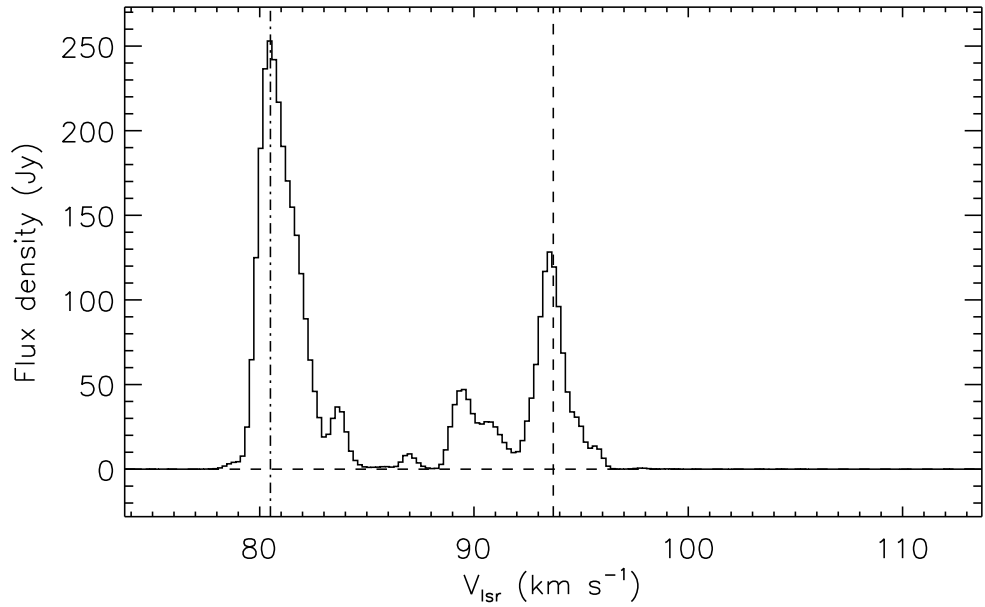}
\caption{Scalar averaged cross-power spectrum of the 22 GHz \water\ maser line toward G25.82. 
The vertical dashed line and the horizontal dashed line represent the systemic velocity of the source and zero baseline, respectively.
The dash-dotted line indicates the velocity channel used as a reference for phase referencing analysis (triangle in \autoref{sio-outflow}). }
\label{water-spec}
\end{center}
\end{figure}

\begin{figure}
\begin{center}
\plotone{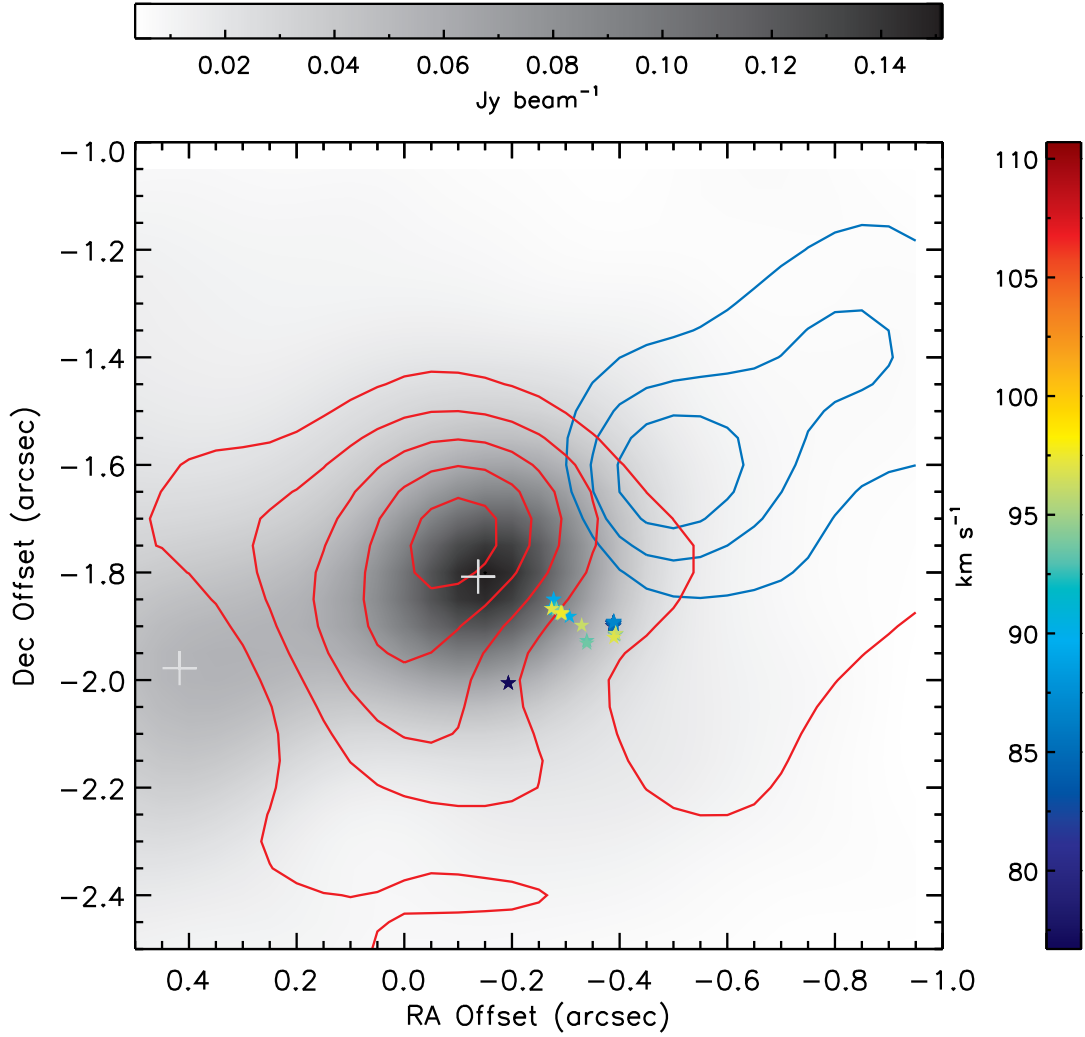}
\caption{Distribution of the 22 GHz \water\ masers obtained with KaVA overlaid onto ALMA 1.3 mm continuum emission (gray scale).
Blue and red contours indicate the integrated intensity map of the \sio\ line as presented in \autoref{sio-outflow}.
Blue contours range from 5$\sigma$ to 15$\sigma$ having the intervals of 5$\sigma$ with the 1$\sigma$ level of 9.2 \inten\ while red contours range from 5$\sigma$ to 25$\sigma$ having the same intervals with the 1$\sigma$ level of 16 \inten.
Stars denote \water\ maser spots having the same velocities in the color bar.
The bluest feature at 41.5 \vel\ at (-0.2, -2.0) is out of range of the color bar.
Crosses denote the positions of the 1.3 mm continuum sources, G25.82--W1 and G25.82--W2.
The absolute position errors of water maser features are in all cases $<$ 1 mas.
In addition, that of ALMA Cycle 3 observations is less than 0.01\arcsec.}
\label{sio-water}
\end{center}
\end{figure}

Several \water\ maser features are identified toward G25.82--W1.
The 22 GHz \water\ maser is often detected in the innermost ($<1000$ au) regions of shocked outflowing gas from the central protostar \citep[e.g.,][]{goddi07,chibueze14,moscadelli16,burns17}. 
Hereafter we call the outflow traced by these \water\ masers the ``\water\ maser outflow''.
The velocity range of detected maser features is from 41.5 \vel\ to 98 \vel\ (\autoref{water-spec}). 
The absolute position of the 80.5 \vel\ feature determined by VERA is $(\alpha, \delta)$=(18$^h$39$^m$03$^s$.60, -06{\degr}24{\arcmin}11{\arcsec}.4) which is also used to astrometrically register the KaVA maser map (\autoref{sio-water}).
The positional errors of water maser features determined using VERA astrometry data are, in all cases, less than 1 mas.
A compact distribution is shown in the southwest part of G25.82--W1 with an offset of 0.25$\arcsec$ ($\sim$1200 au). 
Most of the maser spots are distributed within a 0.2\arcsec\ area except one spot at 41.5 \vel, which is located 0.2$\arcsec$ southeast from the 80.5 \vel\ component. 
Within the 0.2$\arcsec$ area, the most red-shifted component (98.0 \vel) is located closest to the position of G25.82--W1 while the most blue-shifted component (77.3 \vel) is located on the southwestern side. 
The velocity structure of the \water\ maser outflow is not consistent with those of other outflow tracers.

\subsubsection{\thmet }
\begin{figure*}
\begin{center}
\includegraphics[angle=-90,scale=0.7]{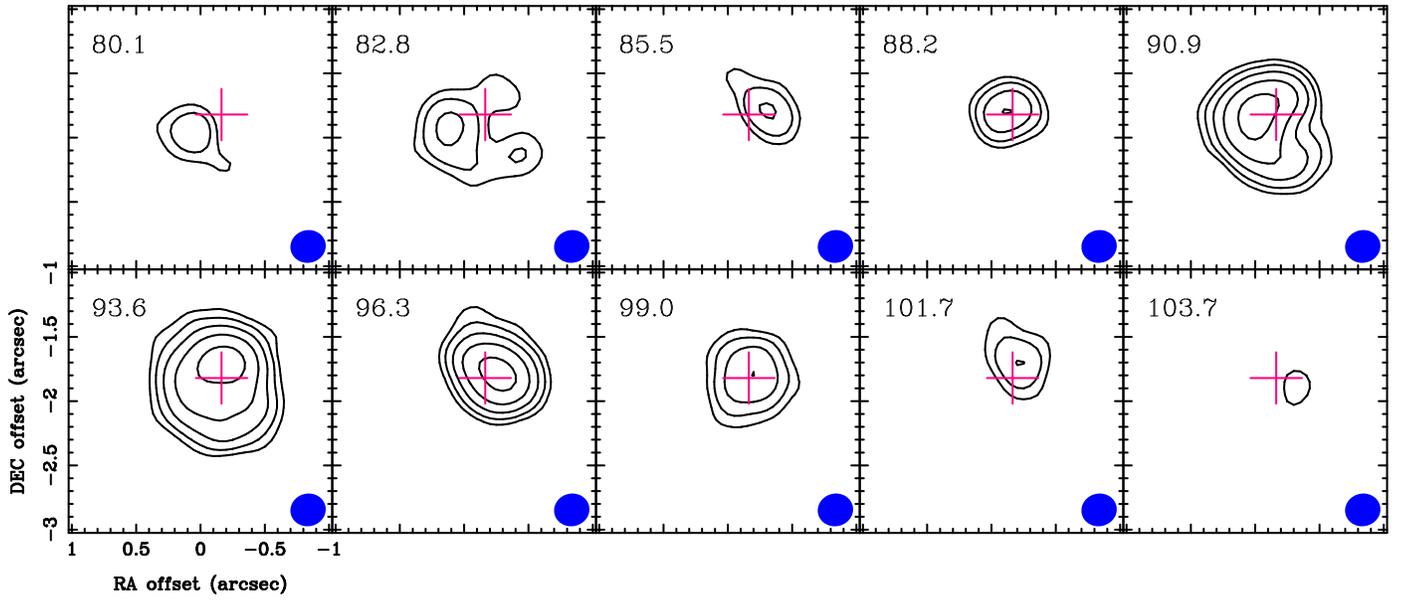}
\caption{Channel maps of the \thmet\ line. 
Respective velocities in \vel\ are presented in the upper left of each panel.
Contours show the 5$\sigma$, 10$\sigma$, 20$\sigma$, 40$\sigma$, and 80$\sigma$ levels, where 1$\sigma$ is 1.24 \mjypb. 
Cross denotes the position of G25.82--W1.
The filled blue circle in the right bottom corner presents the synthesized beam size in the \thmet\ line map.}
\label{ch3oh-chan}
\end{center}
\end{figure*}

\begin{figure}
\begin{center}
\plotone{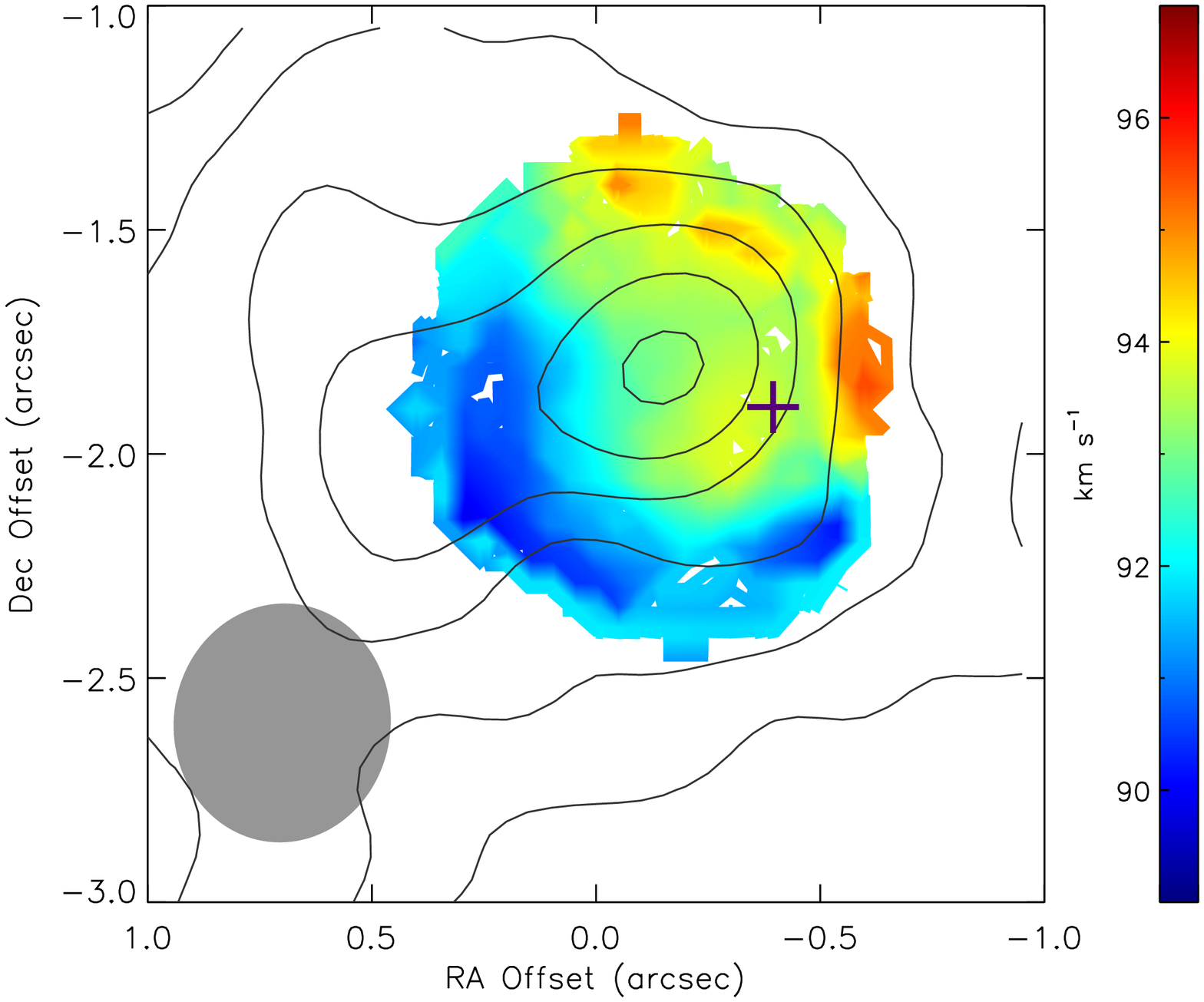}
\caption{The moment 1 map of the \thmet\ line (color scale). 
Contours show the distribution of the 1.3 mm continuum emission.
Contour levels are 10$\sigma$, 20$\sigma$, 40$\sigma$, 80$\sigma$, 160$\sigma$, and 320$\sigma$, where 1$\sigma$ is 0.4 \mjypb.
A purple cross denotes the absolute position of the \water\ maser feature at 80.5 \vel.
The filled gray circle in the bottom left corner indicates synthesized beam size.
The absolute position errors of water maser features are in all cases $<$ 1 mas.
In addition, that of ALMA Cycle 3 observations is less than 0.01\arcsec.}
\label{ch3oh-mom1}
\end{center}
\end{figure}

Channel maps of \thmet\ line emission are shown in \autoref{ch3oh-chan}. 
The 230 GHz \met\ line emission shows a compact ($\sim1\arcsec$) circular distribution at the continuum peak of G25.82--W1. 
There are no emission peaks at G25.82--W2 or W3.
However, the peak positions of individual channels are not exactly coincident with that of the 1.3 mm continuum of G25.82--W1.
This ring-like structure with a velocity gradient can be found in the intensity-weighted mean velocity (moment 1) map as shown in \autoref{ch3oh-mom1}.      
The velocity gradient ranges ranges from 90.2 \vel\ to 96.1 \vel\ within a diameter of about 1\arcsec, which corresponds to 5000 au. 
Overall, the 230 GHz \met\ line emission appears to trace a rotating disk or envelope around G25.82--W1.

\section{Discussions}
\subsection{Mass Estimation}
Because there is no infrared or centimeter radio continuum source in G25.82--W, we assume all the continuum emission at 1.3 mm is thermal dust emission.
Assuming the emission is optically thin graybody radiation, we can derive the total enclosed mass (gas+dust), $M_{\textrm{total}}$, using the following equation:
\begin{equation}
M_{\textrm{total}} = \frac{S_{\nu}{D}^{2}}{{\kappa}_{\nu}B_{\nu}(T_{\textrm{dust}})}.
\end{equation}
The 1.3 mm integrated flux, S$_{\nu}$, in Table 1 for G25.82--W1 (377 mJy) is used and $D$ is the adopted distance of 5.0 kpc.
Since the millimeter/submillimeter continuum data at other wavelengths is not available for this source, dust temperature ($T_{\textrm{dust}}$) is assumed to be 150 K with an uncertainty by a factor of 2 (75 K - 300 K).
The $B_{\nu}$ is the Planck function for a blackbody of temperature $T_{\textrm{dust}}$.
For the mass opacity coefficient, we adopted $\kappa_{\nu}=0.1(250 \mu m/{\lambda})^{\beta}$ cm$^{2}$g$^{-1}$ from \citet{hildebrand83} with $\beta=1.85$ from the dust model of \citet{ossenkopf94}.
The total mass is estimated to be 20 M$_\odot$-84 M$_\odot$ assuming the gas to dust mass ratio of 100.
The luminosity can also be estimated using the above assumptions and the following relation \citep{scoville76, zhang07};
\begin{equation}
T_{\textrm{dust}} = 65\left (\frac{0.1\ \textrm{pc}}{R_{\textrm{s}}}\right )^{2/(4+\beta)}\left (\frac{L}{10^5\ L_{\odot}}\right )^{1/(1+\beta)}\left (\frac{0.1}{\kappa_{\nu}}\right )^{1/(4+\beta)} K.
\end{equation}
Here $R_{\textrm{s}}$ is the source radius in pc. 
Using a source size of 0.005 pc (0.22\arcsec), the luminosity of G25.82--W1 is estimated to be $\sim1.3\times10^{4}$ L$_{\odot}$ corresponding to a B0.5 star having a stellar mass of 12.7 M$_{\odot}$ \citep{mottram11}.

As for G25.82--W2 and W3, there is no molecular emission peak at any of them suggesting that there is no heating source (or embedded YSO) in each of them.
Therefore, we here assume that they are starless cores.
The derived masses of these cores are 361 M$_{\odot}$ and 99 M$_{\odot}$, respectively, adopting a 15 K dust temperature for starless massive dense cores \citep{russeil10}.

Dynamical mass of the central protostar can be estimated assuming that the 230 GHz \met\ line emission arises from an edge-on rotating structure with the Keplerian rotation,
\begin{equation}
M=\frac{Rv^2}{G},
\end{equation}
where $G$ is the gravitational constant, $R$ is the radius of the 230 GHz \thmet\ line emission of 2500 au (0.5\arcsec) and $v$ is the velocity difference within $R$ of 3 \vel. 
The derived mass is 25 $\pm$ 11 M$_\odot$ in the case of the edge-on geometry.
The uncertainty in dynamical mass includes the fractional errors of the measured radius mainly caused by the uncertainties in the kinematic distance (6$\%$) and the velocity difference of 0.65 \vel (23$\%$).  
The measurement error in defining the size of radius is small compare to that in the kinematic distance.
Our mass estimate is a lower limit because the inclination of the disk is unknown.
Thus, G25.82--W1 is very likely a high-mass protostellar object.

\subsection{Evolutionary phases of G25.82--E and G25.82--W}
G25.82 is composed of two regions, G25.82--E and G25.82--W. 
G25.82--E is the weakest 1.3 mm continuum condensation in G25.82.
There is one WISE 22 $\micron$ source, J183903.80-062411.0, in the vicinity of G25.82--E.
In addition, a radio continuum source was also detected in this region using the VLA \citep{hu16}.
Considering the spatial resolution of the VLA observations ($\sim3\arcsec$), the radio continuum source appears to be associated with G25.82--E.
In the ultra-compact H\textsc{ii} (UCH\textsc{ii}) region phase, the free-free radiation is the dominant source of the centimeter radio continuum emission from central HM-YSOs.
Spatial association with far-infrared (FIR) and centimeter radio continuum sources suggests that G25.82--E is at an evolved stage such as an UCH\textsc{ii} region.
The weak 1.3 mm continuum emission of G25.82--E suggests the destruction of dust grains by UV radiation from the central HM-YSO.  
Furthermore, there is no signpost of mass accretion traced by molecular lines such as outflows or rotating/accreting material around the HM-YSO.

On the other hand, G25.82--W is an infrared dark region.
Even at the WISE 22 \micron, there is no counterpart.
Moreover, radio continuum emission is not detected toward this region \citep{hu16}.
However, multiple outflows in different scales are identified in the present study.
The N--S SiO outflow, the SE--NW SiO outflow, and \water\ maser outflow are driven by YSOs in G25.82--W.
Outflows are strong evidence that the accretion process is ongoing in this region. 
Therefore it follows that G25.82--W is in an earlier evolutionary stage than G25.82--E.

\subsection{Outflows in G25.82--W}

Within G25.82--W, there are at least three potential sites of star formation; an HM-YSO, G25.82--W1, and two starless cores, G25.82--W2, and G25.82--W3.
The size of this region is determined by the 10$\sigma$ contour level in the 1.3 mm continuum map to be about 2$\arcsec$ or 0.05 pc. 
This is smaller than the typical size of massive dense cores, 0.1-0.2 pc \citep{motte17}.
Clustered formation in this small region is also suggested by the existence of multiple outflows.

Here we discuss that the N--S SiO outflow is most likely driven by G25.82--W1 as it is located at the center of blue- and red-shifted components.
In addition, a rotating structure is also detected toward G25.82--W1.
The velocity gradient shown in the moment 1 map of the \thmet\ line traces a rotating disk or envelope around G25.82--W1.
Therefore, it is likely that G25.82--W1 hosts an embedded HM-YSO.

In the case of the SE--NW SiO outflow, the driving source cannot be determined based on the morphology of SiO outflows and the distribution of continuum sources at the resolution of the ALMA data presented here.
Moreover, no signs of star formation in G25.82--W2 and W3 suggests that they have less possibility to be the driving source of the SE--NW SiO outflow.
In addition, the existence of a potential companion is suggested by the spatial inconsistency between the \water\ maser and the 1.3 mm continuum peak if the \water\ maser outflow were to be tracing another small scale outflow not associated with the SiO outflows.
However, we cannot rule out the possibility that the \water\ maser outflow is tracing the inner part of the large scale outflow (N--S SiO outflow). 
If this is the case, the inconsistency of velocity structure of \water\ maser outflow and N--S SiO outflow could be interpreted by the difference in physical structures traced by each tracer. 
In other word, the \water\ maser outflow may be the interior of outflowing gas in the vicinity of G25.82-W1.

There are potential HM-YSOs in G25.82--W, G25.82--W1 and a possible companion hosting the SE--NW SiO outflow. 
In addition, high-mass starless$/$protostellar core candidates are also detected, G25.82--W2 and G25.82--W3.
In contrast, low-mass starless$/$protostellar cores are not detected considering the sensitivity of continuum emission at 1.3 mm (5$\sigma\sim$2 \mjypb$\sim$0.5 M$_{\odot}$) which is high enough to detect low-mass cores.
This implies that these HM-YSOs formed in isolation as predicted by the turbulent core accretion model (e.g., CygX-N53-mm2 \citet{duarte13}, G11.92-0.61-MM2 \citet{cyganowski14}, and C9A \citet{kong17}). 
Therefore, high-mass star formation in G25.82 can be a scaled-up version of low-mass star formation where the material accretes onto the YSOs through an accretion disk within a monolithic core.
Our results do not show the presence of a cluster of low-mass cores around the detected three high-mass cores, although we cannot exclude the possibility that there are embedded low-mass cores in G25.82--W which were not resolved in our observations.
This work promotes the G25.82 region as a valuable laboratory for future investigations of high-mass star formation.

\section{Summary}
To investigate the mechanism of high-mass star formation in the G25.82 region, we carried out 1.3 mm continuum and spectral line observations with ALMA and 22 GHz \water\ maser observations using KaVA and VERA. 
Our main results are summarized as follows:

\begin{enumerate}
\item{In the 1.3 mm continuum map, G25.82 is divided into two parts; G25.82--E and G25.82--W. 
G25.82--E is a compact and weak condensation being located close to a source of infrared and radio continuum.
On the other hand, no infrared or radio continuum source is detected toward the other strong 1.3 mm continuum source, G25.82--W.  
Three 1.3 mm continuum condensations are identified, G25.82--W1, W2, and W3.
The strongest 1.3 mm continuum peak is located at G25.82--W1.}

\item{ In G25.82--W, two SiO bipolar outflows are identified.
One is relatively poorly collimated outflow having blue- and redshifted components at north and south, respectively (N--S SiO outflow).
Both blue- and redshifted components show curved structures. 
The other shows weaker emission but still significant ($>5\sigma$), extending in the southeast--northwest direction (SE--NW SiO outflow).
Blue-shifted knots are elongated to the SE of G25.82--W while red-shifted knots are located at the NW. 
The axis of the SE--NW SiO outflow is shifted at least $0.8\arcsec$ to the southwest from the 1.3 mm continuum sources.}

\item{ The \masmeto\ line at 229 GHz is also detected toward G25.82--W.
The distribution of the \masmeto\ line shows a mixture of thermal and maser emission. 
The extended structure ($\sim15\arcsec$) close to the systemic velocity shows a similar structure to that of the \sio\ line emission in the same velocity ranges. }

\item{In contrast, the 22 GHz \water\ masers are distributed at a position 0.25$\arcsec$ shifted from that of G25.82--W1.
The velocity distribution of the \water\ masers is not consistent with those of other outflow tracers.}

\item{ A ring-like structure is detected around G25.82--W1 ($R\sim$2500 au) in the 230 GHz \thmet\ line emission map. 
The velocity gradient is detected centered at the position of the 1.3 mm continuum peak of G25.82--W1.
It is likely tracing a rotating disk or envelope around a HM-YSO. }

\item{ The total mass of G25.82--W1 is estimated as 20$_\odot$-84$_\odot$ using the 1.3 mm continuum emission while the dynamical mass of $>$ 25 M$_\odot$ is measured using the \thmet\ emission line.
The dynamical mass gives the lower limit due to the inclination of the rotation axis.
Thus, G25.82--W1 appears to be a HM-YSO surrounded by a rotating disk or envelope.
In addition, the luminosity of G25.82--W1, $\sim1.2\times10^{4}$ L$_{\odot}$, also suggests that G25.82--W1 is a high-mass B0.5 type star with a stellar mass of 12.7 M$_\odot$. }

\item{ Given the multiple outflows and multiple sources, G25.82 is a high-mass cluster formation.
G25.82--E is in the UCH\textsc{ii} phase while G25.82--W is in an earlier evolutionary phase due to the lack of infrared or cm radio continuum counterparts.
Signposts of star formation such as outflowing gas and a rotating structure are shown toward G25.82--W supporting that it is in the early evolutionary stage of high-mass star formation.}

\item{Velocity and spatial inconsistency are shown between each outflow map traced by different transitions. 
The driving source of the N--S SiO outflow is G25.82--W1 while that of the SE--NW SiO outflow is uncertain. 
In the case of the \water\ maser outflow, it could be tracing the interior of the larger scale of outflow (N--S SiO outflow) near the central source, G25.82--W1.  }

\item{ Multiple high-mass starless$/$protostellar core candidates are detected in G25.82.
In contrast, low-mass cores were not detected around G25.82--W despite the high sensitivity of our observations.
It implies that HM-YSOs here would be formed in individual high-mass cores, as predicted by the turbulent core accretion model.
Although we cannot fully rule out a possibility of unresolved embedded low-mass cores within G25.82--W, high-mass star formation processes in G25.82 could be consistent with a scaled-up version of low-mass star formation model.} \\

Further studies with higher spatial and spectral resolutions could reveal the detailed dynamics of the disk--outflow systems in G25.82--W. 
As future work, monitoring observations of \water\ masers with KaVA are in progress to investigate the 3D velocity distribution by measuring the proper motions of \water\ maser features.
These result will provide important clues to constrain the direction and inclination of the outflow.  
By comparing the high-velocity outflow (the \sio\ line) and water masers, the relation between outflows at different scales and their powering sources could be investigated. 

\end{enumerate}

This paper makes use of the following ALMA data: ADS/JAO.ALMA$\#$2015.1.01288.S.
ALMA is a partnership of ESO (representing its member states), NSF (USA) and NINS (Japan), together with NRC (Canada), NSC and ASIAA (Taiwan), and KASI (Republic of Korea), in cooperation with the Republic of Chile. 
The Joint ALMA Observatory is operated by ESO, AUI/NRAO, and NAOJ. 
Data analysis was in part carried out on the Multi-wavelength Data Analysis System operated by the Astronomy Data Center (ADC), National Astronomical Observatory of Japan.
T.H. is financially supported by the MEXT/JSPS KAKENHI Grant Number 17K05398.
R.B. acknowledges support through the EACOA Fellowship from the East Asian Core Observatories Association.

\software{\textit{Miriad} \citep{sault95}, CASA \citep[v4.7;][]{mcmullin07}, AIPS \citep{vanmoorsel96}}

\begin{deluxetable}{lcccccc}
\tabletypesize{\footnotesize}
\tablecaption{Summary of observed emissions}
\tablehead{	\colhead{} &
        \colhead{Transition} & 
			\colhead{${\nu}$} & 
			\colhead{${E_{L}}$} &
			\colhead{rms} &
			\colhead{Reference$^{a}$} &
			\colhead{Telescope} \\
			\colhead{} &
			\colhead{} &
			\colhead{[GHz]} & 
			\colhead{K} &
			\colhead{[\mjypb]} &
			\colhead{} &
			\colhead{}}
		
\startdata
\water & 6$_{1,6}$-5$_{2,3}$ & 22.235080 & 642.43 & 30 & JPL & KaVA, VERA\\ 
SiO & 5--4 & 217.104980 & 20.84 &  1.45 & JPL & ALMA\\
\met & 8$_{-1}$--7$_{0}$ E & 229.758811 & 78.08& 1.43 & JPL & ALMA \\
\met & 22$_{4}$--21$_{5}$ E & 230.368199 & 671.69 &  1.24 & JPL & ALMA \\
Continuum &  &  &  & 0.40 & \nodata & ALMA
\enddata
\tablenotetext{a}{Jet Propulsion Laboratory Catalog \citep[JPL;][]{pickett98}.}
\end{deluxetable}

\begin{deluxetable*}{lccccccc}
\tabletypesize{\footnotesize}
\tablecaption{Parameters of the 1.3 mm Continuum }
\tablehead{      \colhead{}&
     \colhead{ RA } &
		    \colhead{ Dec } &
			\colhead{ Peak Flux} &
			\colhead{ Integrated Flux Density} &
			\colhead{ Source Size } \\
			\colhead{} &
			\colhead{[$^{h}$ $^{m}$ $^{s}$]} &
			\colhead{[$\degr$ $\arcmin$ $\arcsec$]} &
			\colhead{[\mjypb]} &
			\colhead{[mJy]} &
			\colhead{[$\arcsec\ \times\ \arcsec$], P.A. $\degr$} }
			
\startdata 
G25.82--W1 & 18:39:03.62 & -6:24:11.31  & 143 (0.2) & 377 (0.8)  &  [0.5 $\times$ 0.4], 116.4\\
G25.82--W2 & 18:39:03.66 & -6:24:11.48 & 53 (0.2) & 237 (1.3) & [0.6 $\times$ 0.5], 178.6\\
G25.82--W3 & 18:39:03.68 & -6:24:12.48 & 13 (0.2) & 65 (1.4)  & [0.7 $\times$ 0.5], 160.5\\
G25.82--E   & 18:39:03.82 & -6:24:11.56 & 8 (0.2) & 31 (1.2) & [0.7 $\times$ 0.5], 80.2
\enddata
\tablenotetext{}{Numbers in parenthesis represent the errors of peak fluxes and flux densities from the Gaussian fitting (1$\sigma$).}

\end{deluxetable*}

\end{document}